\documentclass{article}
\usepackage{amsmath}
\usepackage{amssymb}
\usepackage{epsfig}

\newtheorem{proposition}{Proposition}
\newtheorem{theorem}{Theorem}
\newtheorem{corollary}{Corollary}
\newtheorem{lemma}{Lemma}

\title{On the number of gapped repeats with arbitrary gap}

\author{
        Roman Kolpakov\\
        Lomonosov Moscow State University,\\  
				Dorodnicyn Computing Centre FRC CSC RAS,\\
				Moscow, Russia\\
        Email: foroman@mail.ru
       }
\date{}

\begin{document}

\maketitle

\begin{abstract}
For any functions $f(x)$, $g(x)$ from $\mathbb {N}$ to $\mathbb {R}$
we call repeats $uvu$ such that $g(|u|)\le |v|\le f(|u|)$ as {\it $f,g$-gapped repeats}.
We study the possible number of $f,g$-gapped repeats in words of fixed length~$n$.
For quite weak conditions on $f(x)$, $g(x)$ we obtain an upper bound on this number
which is linear to~$n$.
\end{abstract}

\section{Introduction}

Let $w=w[1]w[2]\ldots w[n]$ be an arbitrary word of length $|w|=n$. 
A fragment $w[i]\cdots w[j]$ of~$w$, where $1\le i\le j\le n$, is called 
a {\it factor} of~$w$ and is denoted by $w[i..j]$. Note that this factor 
can be considered either as a word itself or as the fragment $w[i]\ldots w[j]$ 
of~$w$. So for factors we have two different notions of equality: factors 
can be equal as the same fragment of the word~$w$ or as the same word. 
To avoid this ambiguity, we use two different notations: if two factors 
$u$ and $v$ of~$w$ are the same word (the same fragment of~$w$) we will 
write $u=v$ ($u\equiv v$). For any $i=1,\ldots,n$ the factor $w[1..i]$ 
($w[i..n]$) is called a {\it prefix} (a {\it suffix}) of~$w$. By positions 
in~$w$ we mean the order numbers $1, 2,\ldots ,n$ of letters of the word~$w$. 
For any factor~$v\equiv w[i..j]$ of~$w$ the positions $i$ and $j$ are called 
{\it start position} of~$v$ and {\it end position} of~$v$ and denoted by 
${\rm beg} (v)$ and ${\rm end} (v)$ respectively. For any two factors $u$, $v$ 
of~$w$ the factor $u$ {\it is contained} in~$v$ if ${\rm beg} (v)\le {\rm beg} (u)$ 
and ${\rm end} (u)\le {\rm end} (v)$. If some word $u$ is equal to a factor~$v$ 
of~$w$ then $v$ is called {\it an occurrence} of~$u$ in~$w$.

We denote by $p(w)$ the minimal period of a word $w$ and by $e(w)$ the ratio 
$|w|/p(w)$ which is called the {\it exponent} of~$w$. A word is called {\it primitive} 
if its exponent is not an integer greater than~1. A word is called {\it periodic} 
if its exponent is greater than or equal to~2. Occurrences of periodic words are 
called {\it repetitions}. Repetitions are fundamental objects, due to their primary 
importance in word combinatorics~\cite{Lothaire83} as well as in various applications, 
such as string matching algorithms~\cite{GaliSeiferas83,CrochRytter95}, molecular 
biology~\cite{Gusfield97}, or text compression~\cite{Storer88}. The simplest 
and best known example of repetitions is factors of the form $uu$, where $u$ 
is a nonempty word. Such repetitions are called {\it squares}. A square $uu$
is called {\it primitive} if $u$ is primitive. The questions on the number of squares 
and effective searching of squares in words are well studied in the literature (see, 
e.g.,~\cite{CrochRytter95,Crochemor81,GusfStoye04}).

A repetition in a word is called {\it maximal} if this repetition cannot be extended 
to the left or to the right in the word by at least one letter with preserving its 
minimal period. More precisely, a repetition $r\equiv w[i..j]$ in~$w$ is called {\it maximal}
if it satisfies the following conditions:
\begin{enumerate}
\item if $i>1$, then $w[i-1]\neq w[i-1+p(r)]$,
\item if $j<n$, then $w[j+1-p(r)]\neq w[j+1]$.
\end{enumerate}
Maximal repetitions are usually called {\it runs} in the literature. Since runs contain 
all the other repetitions in a word, the set of all runs can be considered as a compact 
encoding of all repetitions in the word which has many useful applications (see, 
for example,~\cite{Crochetal1}). For any word~$w$ we will denote by ${\rm E}(w)$ the sum 
of exponents of all maximal repetitions in~$w$. The following bound for ${\rm E}(w)$ is 
proved in~\cite{KK00}.
\begin{theorem}
${\rm E}(w)=O(n)$ for any~$w$.
\label{onsumexp}
\end{theorem}
More precise upper bounds on ${\rm E}(w)$ were obtained in~\cite{CrochIlieTinta, Crochetal11, RunsTheor}.

\begin{figure}[tb]
\centerline{
\psfig{figure=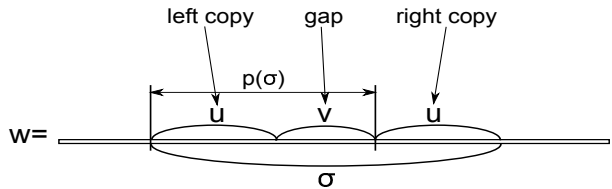,height=24mm,width=80mm}
}
\caption{A gapped repeat $\sigma$ in~$w$.}
 \label{gaprep}
\end{figure}

\begin{figure}[tb]
\centerline{
\psfig{figure=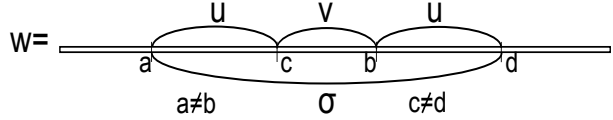,height=15mm,width=80mm}
}
\caption{A maximal gapped repeat $\sigma$ in~$w$.}
 \label{maxgaprep}
\end{figure}

A natural generalization of squares is factors of the form $uvu$ where $u$ and $v$ are
nonempty words. We call such factors {\it gapped repeats}. In the gapped repeat $uvu$
the first (second) factor~$u$ is called {\it the left (right) copy}, and $v$ is called 
{\it the gap}. By {\it the period} of this gapped repeat we will mean the value $|u|+|v|$.
For a gapped repeat~$\sigma$ we denote the length of copies of~$\sigma$ by $c(\sigma)$
and the period of~$\sigma$ by $p(\sigma)$ (see Fig.~\ref{gaprep}).
By $(u', u'')$ we will denote the gapped repeat with the left copy~$u'$ and the right copy~$u''$.
Not that gapped repeats may form the same segment but have different periods. Such repeats are 
considered as distinct, i.e. a gapped repeat is not specified by its start and end positions in 
the word because these positions are not sufficient for determining the both copies and the gap of
the repeat. Analogously to repetitions, a gapped repeat $(w[i'..j'], w[i''..j''])$ in~$w$ is called 
{\it maximal} if it satisfies the following conditions:
\begin{enumerate}
\item if $i'>1$, then $w[i'-1]\neq w[i''-1]$,
\item if $j''<n$, then $w[j'+1]\neq w[j''+1]$.
\end{enumerate}
In other words, a gapped repeat in a word is maximal if its copies cannot be extended 
to the left or to the right in the word by at least one letter with preserving its period 
(see Fig.~\ref{maxgaprep}).

Let $f(x)$, $g(x)$ be functions from $\mathbb {N}$ to $\mathbb {R}$ such
that $0<g(x)\le f(x)$ for any $x\in \mathbb {N}$. We call a gapped repeat
$uvu$ {\it $f,g$-gapped repeat} if $g(|u|)\le |v|\le f(|u|)$. To our knowledge,
maximal $f,g$-gapped repeats were firstly investigated in~\cite{Brodal00} where
it was shown that for computed in constant time functions $f$, $g$ all maximal  
$f,g$-gapped can be found in a word of length~$n$ with time complexity $O(n\log n+S)$ 
where $S$ is the size of output. An algorithm for finding in a word all gapped repeats 
with a fixed gap length in time $O(n\log d+S)$ where $d$ is the gap length, $n$ is the 
word length, and  $S$ is the size of output was proposed in~\cite{KK00a}. 
$f,g$-gapped repeats is a natural generalization of gapped repeats~$\sigma$
such that $p(\sigma)\le\alpha c(\sigma)$ for some $\alpha >1$. Such gapped repeats
which can be considered as a particular case of $f,g$-gapped repeats for
$f(x)=(\alpha -1)x$ and $g(x)=\min \{1, \alpha -1\}$ are called {\it $\alpha$-gapped repeats}.
The notion of $\alpha$-gapped repeats was introduced in~\cite{KPPHr13} where 
it was proved that the number of maximal $\alpha$-gapped repeats  in a word of 
length~$n$ is bounded by $O(\alpha^2 n)$ and all maximal $\alpha$-gapped repeats 
can be found in $O(\alpha^2 n)$ time for the case of integer alphabet. A new approach 
to computing $\alpha$-gapped repeats was proposed in~\cite{GabrMan} in~\cite{GabrMan} 
where it was shown that the longest $\alpha$-gapped repeat in a word of length~$n$ 
over an integer alphabet can be found in $O(\alpha n)$ time. In~\cite{Tanimuraetal} 
an algorithm using an approach previously introduced in \cite{BadkobehCrochToop12} 
is proposed for finding all maximal $\alpha$-gapped repeats in $O(\alpha n+S)$ time 
where $S$ is the output size, for a constant-size alphabet. Finally, in~\cite{LATA16,GawrychowskiIIK16}
an asymptotically tight $O(\alpha n)$ bound on the number of maximal $\alpha$-gapped repeats 
in a word of length~$n$ was independently proved and, moreover, algorithms for finding of
all maximal $\alpha$-gapped repeats in $O(\alpha n)$ time were proposed.

For any real~$x$ denote
$$
|x|^+=\begin{cases}
x, \text{ if } x>0;\\
0, \text{ otherwise; }
\end{cases}\quad
|x|^-=\begin{cases}
-x, \text{ if } x<0;\\
0, \text{ otherwise; }
\end{cases}
$$
Let $f(x)$ be a function from ${\bf N}$ to ${\bf R^+}$. For each $x\in {\bf N}$
denote $\partial^+_f(x)=|f(x+1)-f(x)|^+$ and $\partial^-_f(x)=|f(x+1)-f(x)|^-$.
Denote also $\sup_{x}\{\partial^+_f(x)\}$ ($\sup_{x}\{\partial^-_f(x)\}$) by
$\partial^+_f$ ($\partial^-_f$) if this supremum exists. Let $f(x)$, $g(x)$
be two function from ${\bf N}$ to ${\bf R^+}$ such that $f(x)\ge g(x)$ for
any $x\in {\bf N}$. If the both values $\partial^+_f$ and $\partial^-_g$
exist denote $\max\{\partial^+_f, \partial^-_g\}$ by $\partial^a_{f, g}$.
If the both values $\partial^-_f$ and $\partial^+_g$ exist denote 
$\max\{\partial^-_f, \partial^+_g\}$ by $\partial^b_{f, g}$. Let either
$\partial^a_{f, g}$ or $\partial^b_{f, g}$ exist. Then we define 
$\partial_{f, g}$ as $\min \{\partial^a_{f, g}, \partial^b_{f, g}\}$ 
if the both values $\partial^a_{f, g}$, $\partial^b_{f, g}$ exist;
otherwise we define $\partial_{f, g}$ as the existing one from the
values $\partial^a_{f, g}$, $\partial^b_{f, g}$. Denote also $\Delta_{f, g}(x)=
\frac{1}{x}(f(x)-g(x))\ge 0$ for each $x\in {\bf N}$ and $\Delta_{f, g}=
\sup_x\{\Delta_{f, g}(x)\}$ if this supremum exists. In the paper we
prove $O(n(1+\max\{\partial_{f, g}, \Delta_{f, g}\}))$ bound on the 
number of maximal $f,g$-gapped repeats in a word of length~$n$.

\section{Auxiliary definitions and results}

Further we will consider an arbitrary word $w=w[1]w[2]\ldots w[n]$ of length~$n$.
We will use the following quite evident fact on maximal repetitions (see, e.g.,~\cite{Lothaire05}[Lemma 8.1.3]).
\begin{lemma}
Two distinct maximal repetitions with the same minimal period~$p$ can not
have an overlap of length greater than or equal to~$p$.
\label{overlap}
\end{lemma}

It is not difficult also to prove this fact (see,e.g,~\cite{forJDA}[Proposition 1]).
\begin{proposition}
If a square $uu$ is primitive, for any two distinct occurrences $v'$
and $v''$ of $uu$ in~$w$ the inequality $|{\rm beg} (v')-{\rm beg} (v'')|\ge |u|$
holds.
\label{onprimsqr}
\end{proposition}

Since any repetition~$r$ contains as a prefix a primitive square with the period $p(r)$,
Proposition~\ref{onprimsqr} easily implies 
\begin{corollary}
For any two distinct occurrences $v'$ and $v''$ of the same repetition~$r$ in~$w$ 
the inequality $|{\rm beg} (v')-{\rm beg} (v'')|\ge p(r)$ holds.
\label{onprimreps}
\end{corollary}

For obtaining our bound on the number of considered repeats, we use the following classification 
of maximal gapped repeats introduced in~\cite{forJDA}. We say that a maximal gapped repeat is 
{\it periodic} if the copies of this repeat are repetitions. The set of all periodic $f,g$-gapped 
repeats in the word~$w$ is denoted by ${\cal PP}_{f,g}$. A maximal gapped repeat is called 
{\it prefix} ({\it suffix}) {\it semiperiodic} if the copies of this repeat are not repetitions, 
but these copies have a periodic prefix (suffix) which length is not less than the half of the 
copies length. The longest periodic prefix in a copy of a prefix semiperiodic repeat is called 
{\it the periodic prefix} of this copy. The set of all prefix (suffix) semiperiodic $f,g$-gapped 
repeats in the word~$w$ is denoted by~${\cal PSP}_{f,g}$ (${\cal SSP}_{f,g}$). A maximal gapped 
repeat is called {\it semiperiodic} if it is either prefix or suffixsemiperiodic. The set of all 
semiperiodic $f,g$-gapped repeats in the word~$w$ is denoted by~${\cal SP}_{f,g}$. Maximal gapped 
repeats which are neither periodic nor semiperiodic are called {\it ordinary}. The set of all 
ordinary $f,g$-gapped repeats in the word~$w$ is denoted by~${\cal OP}_{f,g}$.

\section{Estimation of maximal $f,g$-gapped repeats}

Further we assume that both the values $\partial_{f, g}$, $\Delta_{f, g}$ exist.
First we estimate the number of periodic maximal $f,g$-gapped repeats in~$w$. 
Let $\sigma\equiv (u', u'')$ be a repeat from~${\cal PP}_{f,g}$.  Then the both 
copies $u'$, $u''$ of $\sigma$ are repetitions in~$w$ which are extended 
respectively to some maximal repetitions $r'$, $r''$ with the same minimal
period in~$w$. If $r'$ and $r''$ are the same repetition~$r$ then we call 
$\sigma$ {\it private} repeat. Otherwise $\sigma$ is called {\it non-private}.
The following bound on the number of private repeats is proved in~\cite{forJDA}[Corollary 4].
\begin{proposition}
The number of private repeats in $w$ is $O(n)$.
\label{onprivate}
\end{proposition}
Let $\sigma\equiv (u', u'')$ be a non-private repeat from ${\cal PP}_{f,g}$,
i.e. $r'$ and $r''$ are distinct repetitions. We will say that $\sigma$ is
{\it generated from left} by~$r'$ ({\it generated from right} by~$r''$) if
$|r'|\le |r''|$ ($|r''|\le |r'|$). We will say also that $\sigma$ is
{\it generated} by a repetition~$r$ if $\sigma$ is generated from left
or from right by~$r$. Let $p$ be the minimal period of $r'$ and $r''$.
Note that if ${\rm beg} (r')<{\rm beg} (u')$ and ${\rm beg} (r'')<{\rm beg} (u'')$
then 
$$
w[{\rm beg} (u')-1]=w[{\rm beg} (u')+p-1]=w[{\rm beg} (u'')+p-1]=w[{\rm beg} (u'')-1]
$$
which contradicts that $\sigma$ is maximal. So either ${\rm beg} (r')={\rm beg} (u')$
or ${\rm beg} (r'')={\rm beg} (u'')$. In an analogous way we have that either 
${\rm end} (r')={\rm end} (u')$ or ${\rm end} (r'')={\rm end} (u'')$.
Let $\sigma$ be generated by the repetition $r'$ ($r''$). If ${\rm beg} (r')<{\rm beg} (u')$ 
and ${\rm end} (r')>{\rm end} (u')$ (or ${\rm beg} (r'')<{\rm beg} (u'')$ and ${\rm end} (r'')>
{\rm end} (u'')$) then, using above observations, we have $r''\equiv u''$ ($r'\equiv u'$),
so $|r''|<|r'|$ ($|r'|<|r''|$) which contradicts that $\sigma$ is generated by $r'$ ($r''$).
Thus the only three following cases are possible.
\begin{enumerate}
\item ${\rm beg} (r')={\rm beg} (u')$ and ${\rm end} (r')>{\rm end} (u')$
(${\rm beg} (r'')={\rm beg} (u'')$ and ${\rm end} (r'')>{\rm end} (u'')$);
\label{fstgencase}
\item ${\rm beg} (r')<{\rm beg} (u')$ and ${\rm end} (r')={\rm end} (u')$
(${\rm beg} (r'')<{\rm beg} (u'')$ and ${\rm end} (r'')={\rm end} (u'')$);
\label{scdgencase}
\item $r'\equiv u'$ ($r''\equiv u''$).
\label{thdgencase}
\end{enumerate}
We will say that $\sigma$ is {\it prefixly generated} by $r'$ ($r''$) in case~\ref{fstgencase},
{\it suffixly generated} by $r'$ ($r''$) in case~\ref{scdgencase}, and
{\it totally generated} by $r'$ ($r''$) in case~\ref{thdgencase}. We denote
the sets of all prefixly generated, suffixly generated and totally generated
repeats from ${\cal PP}_{f,g}$ by ${\cal PPP}_{f,g}$, ${\cal SPP}_{f,g}$ and
${\cal TPP}_{f,g}$ respectively. Thus, any non-private repeat from ${\cal PP}_{f,g}$
belongs to one of the sets ${\cal PPP}_{f,g}$, ${\cal SPP}_{f,g}$, ${\cal TPP}_{f,g}$.
We estimate separately the numbers of repeats in these sets.

\begin{lemma} 
Any maximal repetition~$r$ in~$w$ generates $O(1+e(r)\Delta_{f, g})$
repeates from ${\cal TPP}_{f,g}$.
\label{genTPP}
\end{lemma} 

{\bf Proof.} 
Note that for any two repeats $\sigma_1\equiv (r, u''_1)$, $\sigma_2\equiv (r, u''_2)$
from ${\cal TPP}_{f,g}$ which are totally generated by~$r$ from right the restrictions
$$
{\rm end} (r)+g(|r|)+1\le {\rm beg} (u''_1), {\rm beg} (u''_2)\le {\rm end} (r)+f(|r|)+1
$$
hold. Moreover, by Corollary~\ref{onprimreps}, we have $|{\rm beg} (u''_1)-{\rm beg} (u''_2)|\ge p(r)$.
Thus, the number of such repeats can not be greater than 
$$
1+\frac{f(|r|)-g(|r|)}{p(r)}\le 1+\frac{|r|\Delta_{f, g}(|r|)}{p(r)}=1+e(r)\Delta_{f, g}(|r|)\le 1+e(r)\Delta_{f, g}.
$$
By analogous way we obtain that the number of repeats from ${\cal TPP}_{f,g}$ which are 
totally generated by~$r$ from left is also not greater than $1+e(r)\Delta_{f, g}$, so
$r$ generates no more than $2(1+e(r)\Delta_{f, g})$ repeats from ${\cal TPP}_{f,g}$.

From Lemma~\ref{genTPP}, using Theorem~\ref{onsumexp}, we obtain the following
bound on $|{\cal TPP}_{f,g}|$.

\begin{corollary}
$|{\cal TPP}_{f,g}|=O(n(1+\Delta_{f, g}))$.
\label{onTPP}
\end{corollary}

Now we estimate $|{\cal PPP}_{f,g}|$. Let $r$ be a repetition in~$w$ which
prefixly generates some repeat~$\sigma$ from ${\cal PPP}_{f,g}$, i.e. one
of copies of~$\sigma$ is contained in~$r$, and the other copy is contained 
in another repetition~$r'$ with the same minimal period. Father we will say
that $\sigma$ is generated by~$r$ with the repetition~$r'$.

\begin{proposition}
Let $r, r'$ be two repetitions in~$w$. Then $r$ prefixly generates with~$r'$
less than $e(r)$ repeats.
\label{proponpref}
\end{proposition}

{\bf Proof.} We consider the case ${\rm beg} (r)<{\rm beg} (r')$, i.e. 
the case of repeats generated by~$r$ with~$r'$ from left (the case 
${\rm beg} (r')<{\rm beg} (r)$ is considered analogously). Let
$\sigma_1\equiv (u_1, u'_1)$, $\sigma_2\equiv (u_2, u'_2)$
be two such repeats, i.e. $u_1$, $u_2$ are prefixes of~$r$ and
$u'_1$, $u'_2$ are suffixes of~$r$. Note that in this case repeats
$\sigma_1$, $\sigma_2$ are uniquely defined by the respective positions
${\rm end} (u_1)$, ${\rm end} (u_2)$. Since $u_1$, $u_2$ are repetitions
with the minimal period $p(r)$, i.e. $|u_1|, |u_2|\ge 2p(r)$, we note that
\begin{eqnarray*}
w[{\rm end} (u_1)-2p(r)+1..{\rm end} (u_1)]&=&w[{\rm end} (r')-2p(r)+1..{\rm end} (r')]\\
&=&w[{\rm end} (u_2)-2p(r)+1..{\rm end} (u_2)],
\end{eqnarray*}
i.e. $w[{\rm end} (u_1)-2p(r)+1..{\rm end} (u_1)]$ and $w[{\rm end} (u_2)-2p(r)+1..{\rm end} (u_2)]$
are equal primitive squares with the period $p(r)$. Hence, by Proposition~\ref{onprimsqr},
we have $|{\rm end} (u_1)-{\rm end} (u_2)|\ge p(r)$. Moreover, since $u_1$, $u_2$ are prefixes of~$r$,
the restrictions
$$
{\rm beg} (r)+2p(r)-1\le {\rm end} (u_1), {\rm end} (u_2)\le {\rm end} (r)
$$
hold. Thus the number of considered repeats is bounded by
$$
1+\frac{|r|-2p(r)}{p(r)}<\frac{|r|}{p(r)}=e(r).
$$

Let $\sigma\equiv (u', u'')$ be a repeat from ${\cal PPP}_{f,g}$ prefixly generated
from left by a maximal repetition~$r$, and $v$ be the gap of~$\sigma$.
Note that 
$$
{\rm beg} (u'')={\rm beg} (u')+|u'|+|v|={\rm beg} (r)+|u'|+|v|.
$$
So, since $g(|u'|)\le |v|\le f(|u'|)$, we have
$$
{\rm beg} (r)+|u'|+g(|u'|)\le {\rm beg} (u'')\le {\rm beg} (r)+|u'|+f(|u'|).
$$
Thus, taking into account $2p(r)\le |u'|<|r|$,  we obtain
$$
{\rm beg} (r)+\min_{2p(r)\le x<|r|}(x+g(x)) \le {\rm beg} (u'')\le
{\rm beg} (r)+\max_{2p(r)\le x<|r|}(x+f(x)).
$$
We conlude this fact, denoting ${\rm beg} (r)+\min_{2p(r)\le x<|r|}(x+g(x))$
by ${\rm lb}_g (r)$ and ${\rm beg} (r)+\max_{2p(r)\le x<|r|}(x+f(x))$ by
${\rm ub}_f (r)$.

\begin{proposition}
Let $\sigma\equiv (u', u'')$ be a repeat from~${\cal PPP}_{f,g}$ prefixly generated
from left by a maximal repetition~$r$. Then
$$
{\rm lb}_g (r)\le {\rm beg} (u'')\le {\rm ub}_f (r).
$$
\label{pronbegu}
\end{proposition}

For any maximal repetition~$r$ in~$w$ denote by ${\cal GPR}_{f,g} (r)$ the set
of all maximal repetitions~$r'$ such that $r$ prefixly generates from left
with~$r'$ at least one repeat from~${\cal PPP}_{f,g}$. Note that all
repetitions from ${\cal GPR}_{f,g} (r)$ have the minimal period $p(r)$.
\begin{proposition}
Let $r$ be a maximal repetition in~$w$. Then for any repetition~$r'$ from 
${\cal GPR}_{f,g} (r)$ except, perhaps, one repetition the conditions
$$
{\rm lb}_g (r)\le {\rm beg} (r')\le {\rm ub}_f (r)
$$
hold.
\label{prongwr}
\end{proposition}

{\bf Proof.} Let $r'$ be a repetition from ${\cal GPR}_{f,g} (r)$, and
$\sigma\equiv (u, u')$ be a repeat prefixly generated from left by~$r$ 
with~$r'$. By Proposition~\ref{pronbegu} we obtain ${\rm beg} (u')\le {\rm ub}_f (r)$,
so ${\rm beg} (r')\le {\rm ub}_f (r)$ since $u'$ is contained in~$r'$.
Thus for any repetition~$r'$ from ${\cal GPR}_{f,g} (r)$ we have
${\rm beg} (r')\le {\rm ub}_f (r)$. Now let $r'_1$, $r'_2$ be two different
repetitions from ${\cal GPR}_{f,g} (r)$ such that 
${\rm beg} (r'_1), {\rm beg} (r'_2)<{\rm lb}_g (r)$, and $\sigma_1\equiv (u_1, u'_1)$,
$\sigma_2\equiv (u_2, u'_2)$ be repeats prefixly generated from left by~$r$ 
with~$r'_1$ and~$r'_2$ respectively. Without loss of generality we assume
that ${\rm end} (u'_1)\le {\rm end} (u'_2)$, so ${\rm end} (u'_1)\le {\rm end} (r'_2)$.
On the other hand, by Proposition~\ref{pronbegu} we have ${\rm lb}_g (r)\le {\rm beg} (u'_1)$,
so ${\rm beg} (r'_2)<{\rm beg} (u'_1)$. Thus $u'_1$ is contained in $r'_2$, so
$u'_1$ is contained in the overlap of $r'_1$ and $r'_2$. Recall that $u'_1$ is
a repetition with the minimal period $p(r)$, so $|u'_1|\ge 2p(r)$. Therefore, $r'_1$ 
and $r'_2$ have the overlap of length greater than $p(r)$ which contradicts
Lemma~\ref{overlap}. Thus no more than one repetition~$r'$ from ${\cal GPR}_{f,g} (r)$
can satisfy the inequality ${\rm beg} (r'_1)<{\rm lb}_g (r)$.

\begin{proposition}
For any maximal repetition~$r$ in~$w$ the bound
$$
{\rm ub}_f (r)-{\rm lb}_g (r)<|r|(1+\Delta_{f, g}+\partial_{f, g})
$$
holds.
\label{ulbound}
\end{proposition}

{\bf Proof.} Let $x^*$ be such that $2p(r)\le x^*<|r|$ and ${\rm beg} (r)+x^*+f(x^*)={\rm ub}_f (r)$,
and $x_*$ be such that $2p(r)\le x^*<|r|$ and ${\rm beg} (r)+x_*+g(x_*)={\rm lb}_g (r)$. 
Then we have
$$
{\rm ub}_f (r)-{\rm lb}_g (r)=(x^*-x_*)+(f(x^*)-g(x_*))<|r|+(f(x^*)-g(x_*)).
$$
Without loss of generality we will assume that $x_*\le x^*$ (the case of $x_*\ge x^*$ 
is considered analogously). We consider separately two possible cases for
the value $\partial_{f, g}$.

a) Let $\partial_{f, g}=\partial^a_{f, g}$. Then
$$
f(x^*)-g(x_*)\le (f(x_*)-g(x_*))+|f(x^*)-f(x_*)|^+
$$
where
$$
f(x_*)-g(x_*)=x_*\cdot\Delta_{f, g}(x_*)<|r|\Delta_{f, g}
$$
and
$$
|f(x^*)-f(x_*)|^+\le \partial^+_f(x^*-x_*)\le \partial^a_{f, g}(x^*-x_*)
<|r|\partial^a_{f, g}=|r|\partial_{f, g}.
$$
Thus $f(x^*)-g(x_*)<|r|(\Delta_{f, g}+\partial_{f, g})$, so
$$
{\rm ub}_f (r)-{\rm lb}_g (r)<|r|(1+\Delta_{f, g}+\partial_{f, g}).
$$

b) Let $\partial_{f, g}=\partial^b_{f, g}$. Then
$$
f(x^*)-g(x_*)\le (f(x^*)-g(x^*))+|g(x^*)-g(x_*)|^+
$$
where
$$
f(x^*)-g(x^*)=x^*\cdot\Delta_{f, g}(x^*)<|r|\Delta_{f, g}
$$
and
$$
|g(x^*)-g(x_*)|^+\le \partial^+_g(x^*-x_*)\le \partial^b_{f, g}(x^*-x_*)
<|r|\partial^b_{f, g}=|r|\partial_{f, g}.
$$
Thus $f(x^*)-g(x_*)<|r|(\Delta_{f, g}+\partial_{f, g})$, so
$$
{\rm ub}_f (r)-{\rm lb}_g (r)<|r|(1+\Delta_{f, g}+\partial_{f, g}).
$$

\begin{corollary}
For any maximal repetition~$r$ in~$w$ the bound 
$|{\cal GPR}_{f,g} (r)|=O(1+\Delta_{f, g}+\partial_{f, g})$ is valid.
\label{gwrbound}
\end{corollary}

{\bf Proof.} Let $r', r''$ be two maximal repetitions from ${\cal GPR}_{f,g} (r)$
such that ${\rm beg} (r')\le {\rm beg} (r'')$. Since $r'$ and $r''$ have the
same minimal period $p(r)$ and $|r'|\ge |r|$ by the definition of ${\cal GPR}_{f,g} (r)$,
by Lemma~\ref{overlap} we note that the overlap of $r'$ and $r''$ is less than $p(r)$, so
$$
{\rm beg} (r'')-{\rm beg} (r')>|r'|-p(r)\ge |r|-p(r)\ge |r|/2.
$$
Therefore, we conclude that the number of repetitions $r'$ from ${\cal GPR}_{f,g} (r)$
such that 
$$
{\rm lb}_g (r)\le {\rm beg} (r')\le {\rm ub}_f (r)
$$
is not greater than $1+\frac{{\rm ub}_f (r)-{\rm lb}_g (r)}{|r|/2}$ which is
less that 
$$
1+\frac{|r|(1+\Delta_{f, g}+\partial_{f, g})}{|r|/2}=O(1+\Delta_{f, g}+\partial_{f, g})
$$
by Proposition~\ref{ulbound}. Thus we obtain that $|{\cal GPR}_{f,g} (r)|=O(1+\Delta_{f, g}+\partial_{f, g})$
by Proposition~\ref{prongwr}.

\begin{lemma}
The number of generated from left repeats from ${\cal PPP}_{f,g}$ is
$O(n(1+\max\{\partial_{f, g}, \Delta_{f, g}\}))$. 
\label{onleftPPP}
\end{lemma}

{\bf Proof.} It follows immediately from Proposition~\ref{proponpref}
and Corollary~\ref{gwrbound} that any maximal repetition~$r$ in~$w$
generates from left $O(e(r)(1+\max\{\partial_{f, g}, \Delta_{f, g}\}))$
repeats from ${\cal PPP}_{f,g}$, so $O(n(1+\max\{\partial_{f, g}, \Delta_{f, g}\}))$
bound in the lemma statement follows from Theorem~\ref{onsumexp}.

In an analogous way we can prove that the number of generated from right 
repeats from ${\cal PPP}_{f,g}$ is also $O(n(1+\max\{\partial_{f, g}, \Delta_{f, g}\}))$. 
So we obtain the following bound on $|{\cal PPP}_{f,g}|$.

\begin{corollary}
$|{\cal PPP}_{f,g}|=O(n(1+\max\{\partial_{f, g}, \Delta_{f, g}\}))$.
\label{onPPP}
\end{corollary}

In an analogous way we can also prove that $|{\cal SPP}_{f,g}|=O(n(1+\max\{\partial_{f, g}, \Delta_{f, g}\}))$.
Thus, using Proposition~\ref{onprivate} and Corollaries~\ref{onTPP} and~\ref{onPPP},
we obtain the following bound on $|{\cal PP}_{f,g}|$.

\begin{corollary}
$|{\cal PP}_{f,g}|=O(n(1+\max\{\partial_{f, g}, \Delta_{f, g}\}))$.
\label{onPP} 
\end{corollary}

Now we estimate the number of semiperiodic maximal $f,g$-gapped repeats in~$w$. 
Let $\sigma\equiv (u', u'')$ be a repeat from~${\cal PSP}_{f,g}$. Denote by $\pi'$ ($\pi''$)
the periodic prefixes of $u'$ ($u''$). Note that these prefixes are extended 
respectively to some distinct maximal repetitions $r'$, $r''$ with the same 
minimal period such that ${\rm end} (\pi')={\rm end} (r')$, ${\rm end} (\pi'')={\rm end} (r'')$.
We will say that $\sigma$ is {\it generated from left} by~$r'$ with~$r''$ 
({\it generated from right} by~$r''$ with~$r'$) if $|r'|\le |r''|$ ($|r''|\le |r'|$).
Let $\sigma$ be generated from left by~$r'$ with~$r''$. Note that
$$
{\rm beg}(u'')={\rm end} (\pi')+1+(|u'|-|\pi'|)+|v|={\rm end} (r')+1+(|u'|-|\pi'|)+|v|,
$$
where $g(|u'|)\le |v|\le f(|u'|)$ and $1\le |u'|-|\pi'|\le |\pi'|\le |r'|$. 
Note also that $1<|u'|\le 2|\pi'|\le 2|r'|$. Therefore,
$$
1+(|u'|-|\pi'|)+|v|\ge 2+g(|u'|)\ge 2+\min_{1< x\le 2|r'|}g(x)
$$
and
$$
1+(|u'|-|\pi'|)+|v|\le 1+|r'|+f(|u'|)\le 1+|r'|+\max_{1< x\le 2|r'|}f(x).
$$
Denoting ${\rm beg} (r')+2+\min_{1< x\le 2|r'|}g(x)$ by ${\rm lp}_g (r')$ 
and ${\rm beg} (r')+1+|r'|+\max_{1< x\le 2|r'|}f(x)$ by ${\rm up}_f (r)$,
we obtain the following fact.
\begin{proposition}
Let $\sigma\equiv (u', u'')$ be a repeat from ${\cal PSP}_{f,g}$ generated
from left by a maximal repetition~$r$. Then
$$
{\rm lp}_g (r)\le {\rm beg} (u'')\le {\rm up}_f (r).
$$
\label{pronbgu}
\end{proposition}

For any maximal repetition~$r$ in~$w$ denote by ${\cal GSR}_{f,g} (r)$ the set
of all maximal repetitions~$r'$ such that $r$ generates from left with~$r'$ at 
least one repeat from~${\cal PSP}_{f,g}$. Note that all repetitions from 
${\cal GSR}_{f,g} (r)$ have the minimal period $p(r)$. Analogously to 
Proposition~\ref{prongwr}, we can prove the following statement.
\begin{proposition}
Let $r$ be a maximal repetition in~$w$. Then for any repetition~$r'$ from 
${\cal GSR}_{f,g} (r)$ except, perhaps, one repetition the conditions
$$
{\rm lp}_g (r)\le {\rm beg} (r')\le {\rm up}_f (r)
$$
hold.
\label{prongsr}
\end{proposition}

\begin{proposition}
For any maximal repetition~$r$ in~$w$ the bound
$$
{\rm up}_f (r)-{\rm lp}_g (r)<|r|(1+2\Delta_{f, g}+2\partial_{f, g})
$$
holds.
\label{ulpound}
\end{proposition}

{\bf Proof.} Let $x^*$, $x_*$, $1<x^*,x_*\le 2|r|$, be such that  $f(x^*)=\max_{1< x\le 2|r'|}f(x)$
and $g(x_*)=\min_{1< x\le 2|r'|}g(x)$.
Then we have
$$
{\rm up}_f (r)-{\rm lp}_g (r)=|r|-1+(f(x^*)-g(x_*))<|r|+(f(x^*)-g(x_*)).
$$
Without loss of generality we will assume that $x_*\le x^*$ (the case of $x_*\ge x^*$ 
is considered analogously). We consider separately two possible cases for
the value $\partial_{f, g}$.

a) Let $\partial_{f, g}=\partial^a_{f, g}$. Then
$$
f(x^*)-g(x_*)\le (f(x_*)-g(x_*))+(f(x^*)-f(x_*))
$$
where
$$
f(x_*)-g(x_*)=x_*\cdot\Delta_{f, g}(x_*)\le 2|r|\Delta_{f, g}
$$
and
$$
f(x^*)-f(x_*)\le \partial^+_f(x^*-x_*)\le \partial^a_{f, g}(x^*-x_*)
<2|r|\partial^a_{f, g}=2|r|\partial_{f, g}.
$$
Thus $f(x^*)-g(x_*)<|r|(2\Delta_{f, g}+2\partial_{f, g})$, so
$$
{\rm up}_f (r)-{\rm lp}_g (r)<|r|(1+2\Delta_{f, g}+2\partial_{f, g}).
$$

b) Let $\partial_{f, g}=\partial^b_{f, g}$. Then
$$
f(x^*)-g(x_*)=(f(x^*)-g(x^*))+(g(x^*)-g(x_*))
$$
where
$$
f(x^*)-g(x^*)=x^*\cdot\Delta_{f, g}(x^*)\le 2|r|\Delta_{f, g}
$$
and
$$
g(x^*)-g(x_*)\le \partial^+_g(x^*-x_*)\le \partial^b_{f, g}(x^*-x_*)
<2|r|\partial^b_{f, g}=2|r|\partial_{f, g}.
$$
Thus $f(x^*)-g(x_*)<|r|(2\Delta_{f, g}+2\partial_{f, g})$, so
$$
{\rm up}_f (r)-{\rm lp}_g (r)<|r|(1+2\Delta_{f, g}+2\partial^b_{f, g}).
$$

Analogously to Corollary~\ref{gwrbound}, from propositions~\ref{prongsr} 
and~\ref{ulpound} we can obtain the following corollary.
\begin{corollary}
For any maximal repetition~$r$ in~$w$ the bound 
$|{\cal GSR}_{f,g} (r)|=O(1+\Delta_{f, g}+\partial_{f, g})$ is valid.
\label{gsrbound}
\end{corollary}

\begin{lemma}
The number of generated from left repeats from ${\cal PSP}_{f,g}$ is
$O(n(1+\max\{\partial_{f, g}, \Delta_{f, g}\}))$. 
\label{onleftPSP}
\end{lemma}

{\bf Proof.} Let $r', r''$ be maximal repetitions in~$w$ such that
$r''$ is contained in ${\cal GSR}_{f,g} (r')$, i.e. there exists
a repeat $\sigma\equiv (u', u'')$ generated from left by~$r'$ with~$r''$.
Let $\pi'$, $\pi''$ be the periodic prefixes of $u'$ and $u''$ respectively,
i.e. $u'\equiv\pi'\nu'$, $u'\equiv\pi''\nu''$ for some suffixes 
$\nu',\nu''$ of $u'$ and $u''$. Since ${\rm end} (\pi')={\rm end} (r')$
and ${\rm end} (\pi'')={\rm end} (r'')$, we can easily see that $\pi'$,
$\pi''$ are the longest common suffixes of $w[1 .. {\rm end} (r')]$ and 
$w[1 .. {\rm end} (r'')]$ respectively, and $\nu',\nu''$ are the 
longest common prefixes of $w[{\rm end} (r')+1 .. n]$ and $w[{\rm end} (r'')+1 .. n]$
respectively. Thus, $\sigma$ is defined uniquely by $r'$ and $r''$, so
$r'$ can generate from left with~$r''$ only one repeat. Therefore, for
any maximal repetition~$r'$ in~$w$ the number of of generated by~$r'$ 
from left repeats from ${\cal PSP}_{f,g}$ is $|{\cal GSR}_{f,g} (r)|$
which is $O(1+\Delta_{f, g}+\partial_{f, g})$ by Corollary~\ref{gsrbound}.
Therefore, $O(n(1+\max\{\partial_{f, g}, \Delta_{f, g}\}))$ bound on the total
number of generated from left repeats from ${\cal PSP}_{f,g}$ follows from
Theorem~\ref{onsumexp}.

In an analogous way we can prove that the number of generated from right
repeats from ${\cal PSP}_{f,g}$ is also $O(n(1+\max\{\partial_{f, g}, \Delta_{f, g}\}))$.
Thus we obtain the same bound on $|{\cal PSP}_{f,g}|$.

\begin{corollary}
$|{\cal PSP}_{f,g}|=O(n(1+\max\{\partial_{f, g}, \Delta_{f, g}\}))$.
\label{onPSP}
\end{corollary}

In an analogous way we can also prove that $|{\cal SSP}_{f,g}|=O(n(1+\max\{\partial_{f, g}, \Delta_{f, g}\}))$.
Thus we obtain the same bound on $|{\cal SP}_{f,g}|$.

\begin{corollary}
$|{\cal SP}_{f,g}|=O(n(1+\max\{\partial_{f, g}, \Delta_{f, g}\}))$.
\label{onSP} 
\end{corollary}

For estimating the number of ordinary maximal $f,g$-gapped repeats in~$w$ 
we use an improvement of approach proposed in~\cite{Kolpakov12, forJDA}.
We consider triples of positive integers $(i, j, c)$. We call such triples 
{\it points}. For any two points $(i', j', c')$, $(i'', j'', c'')$ we say 
that the point $(i', j', c')$ {\it covers from above} ({\it covers from below})
the point $(i'', j'', c'')$ if $i'\le i''\le i'+c'/6$, $j'\le j''\le j'+c'/6$, 
$c'\ge c'' \ge \frac{2c'}{3}$ ($c'\le c'' \le \frac{3c'}{2}$). We represent any 
maximal repeat~$\sigma\equiv (u', u'')$  from ${\cal OP}_{f, g}$ by the point 
$(i, j, c)$ where $i={\rm beg} (u')$, $j={\rm beg} (u')$ and $c=c(\sigma)=|u'|=|u''|$.
It is obvious that $\sigma$ is uniquely defined by the values $i$, $j$ and~$c$, 
so two different repeats from ${\cal OP}_{f, g}$ can not be represented by the
same point. A point is {\it covered from above} ({\it covered from below}) by~$\sigma$ 
if this point is covered from above (covered from below) by the point representing~$\sigma$.
By $V^a[\sigma ]$ ($V^b[\sigma ]$) we denote the set of all points covered from above
(covered from below) by the repeat~$\sigma$. We will call two factors $u_1$, $u_2$ in~$w$ 
{\it strongly overlapped} if either ${\rm beg} (u_1)\le {\rm beg} (u_2)\le {\rm beg} (u_1)+|u_1|/6$ 
or ${\rm beg} (u_2)\le {\rm beg} (u_1)\le {\rm beg} (u_2)+|u_2|/6$. The proof of
the following lemma can be easily deduced from the proof of Lemma~3 in~\cite{LATA16}
but for the sake of clearness we present this proof.

\begin{lemma}
Let $\sigma_1\equiv (u'_1, u''_1)$, $\sigma_2\equiv (u'_2, u''_2)$ be two different 
maximal gapped repeats in~$w$ such that $c(\sigma_1)\le c(\sigma_2)\le\frac{3}{2}c(\sigma_1)$,
the copies $u'_1$, $u'_2$ are strongly overlapped, and the copies $u''_1$, $u''_2$ 
are strongly overlapped. Then either $\sigma_1$ or $\sigma_2$ is not an ordinary repeat.
\label{tehlemma}
\end{lemma}

{\bf Proof.} Denote $c_1=c(\sigma_1)$, $c_2=c(\sigma_2)$, $p_1=p(\sigma_1)$,
$p_2=p(\sigma_2)$, and $d=|p_1-p_2|$. Since $\sigma_1$ and $\sigma_2$ have
overlapped left copies, in the case of $p_1=p_2$ these repeats must coincide 
due to their conditions of maximality. Thus, $p_1\neq p_2$, i.e. $d>0$. Note 
that $p_1={\rm beg}(u''_1)-{\rm beg}(u'_1)$, $p_2={\rm beg}(u''_2)-{\rm beg}(u'_2)$. 
Hence, since both the copies $u'_1$, $u'_2$ and the copies $u''_1$, $u''_2$ are 
strongly overlapped, we have
$$
p_1-|u'_1|/6-|u''_2|/6\le p_2\le p_1+|u'_2|/6+|u''_1|/6,
$$
so $0<d\le (c_1+c_2)/6\le\frac{5c_1}{12}$.

First consider the case when one of the copies $u'_1, u'_2$ is contained in
the other, i.e. $u'_1$ is contained in $u'_2$. In this case, $u''_2$
contains some factor $\widehat u''_1$ corresponding to the factor $u'_1$ in
$u'_2$. Since ${\rm beg}(u''_1)-{\rm beg}(u'_1)=p_1$, ${\rm beg}(\widehat
u''_1)-{\rm beg}(u'_1)=p_2$ and $u''_1=\widehat u''_1=u'_1$, we have
$$
|{\rm beg}(u''_1)-{\rm beg}(\widehat u''_1)|=d,
$$
so $d$ is a period of $u''_1$ such that $d\le\frac{5}{12}c_1=\frac{5}{12}|u''_1|$. 
Thus, $u''_1$ is periodic which contradicts that $\sigma_1$ is not periodic.

Now consider the case when $u'_1, u'_2$ are not contained in one another.
Denote by $z'$ the overlap of  $u'_1$ and $u'_2$. Let $z'$ be a suffix of
$u'_l$ and a prefix of $u'_r$ where $l, r=1, 2$, $l\neq r$. Then $u''_l$
contains a suffix $z''$ corresponding to the suffix $z'$ in $u'_l$, and $u''_r$ 
contains a prefix $\widehat z''$ corresponding to the prefix $z'$ in $u'_r$. 
Since ${\rm beg}(z'')-{\rm beg}(z')=p_l$, ${\rm beg}(\widehat z'')-{\rm beg}(z')=p_r$ 
and $z''=\widehat z''=z'$, we have
$$
|{\rm beg}(z'')-{\rm beg}(\widehat z'')|=|p_l-p_r|=d,
$$
therefore $d$ is a period of $z'$. Note that in this case
$$
{\rm beg}(u'_l)<{\rm beg}(u'_r)\le {\rm beg}(u'_l)+c_l/6,
$$
therefore $0<{\rm beg}(u'_r)-{\rm beg}(u'_l)\le c_l/6$. Thus
$$
|z'|=c_l-({\rm beg}(u'_r)-{\rm beg}(u'_l))\ge\frac{5}{6}c_l\ge\frac{5}{6}c_1.
$$
From $d\le\frac{5}{12}c_1$ and $|z'|\ge\frac{5}{6}c_1$ we obtain $d\le |z'|/2$. 
Thus, $z'$ is a periodic suffix of $u'_l$ such that $|z'|\ge \frac{5}{6}|u'_l|$, 
so $\sigma_l$ is either suffix semiperiodic or periodic, i.e. $\sigma_l$ is not 
an ordinary repeat. 

Note that if two different maximal gapped repeats $\sigma_1\equiv (u'_1, u''_1)$, 
$\sigma_2\equiv (u'_2, u''_2)$ in~$w$ where $c(\sigma_1)\le c(\sigma_2)$ cover
from above or from below the same point then $c(\sigma_2)\le\frac{3}{2}c(\sigma_1)$,
and both the copies $u'_1$, $u'_2$ and the copies $u''_1$, $u''_2$ are strongly 
overlapped. So we can conclude from Lemma~\ref{tehlemma} that any point can not be 
covered from above or from below by two different ordinary repeats.

\begin{corollary}
Two different ordinary repeats in~$w$ can not cover from above the same point.
\label{keycorra}
\end{corollary}
\begin{corollary}
Two different ordinary repeats in~$w$ can not cover from below the same point.
\label{keycorrb}
\end{corollary}

Denote by ${\cal CQ}^a$ (${\cal CQ}^b$) the set of all points $(i, j, c)$ covered from 
above (covered from below) by maximal gapped repeats from ${\cal P}_{f, g}$.

\begin{lemma}
Let both the values $\partial^+_f$, $\partial^-_g$ exist. Then for any point $(i, j, c)$ 
from ${\cal CQ}^a$ the inequalities
$$
i+\frac{5}{6}c+g(c)-\frac{c}{2}\partial^-_g\le j\le i+\frac{7}{4}c+f(c)+\frac{c}{2}\partial^+_f
$$
hold.
\label{coverlemma}
\end{lemma}

{\bf Proof.} Let a point $(i, j, c)$ belongs to ${\cal CQ}^a$, i.e. this point
is covered from above by some point $(i', j', c')$ representing the gapped repeat
$\sigma'\equiv (u', u'')$ from ${\cal P}_{f, g}$ such that $i'={\rm beg} (u')$, 
$j'={\rm beg} (u'')$ and $c'=c(\sigma')$. Thus we have $i'\le i \le i'+c'/6$,
$j'\le j \le j'+c'/6$, and $c'\ge c \ge \frac{2c'}{3}$.
Since $\sigma'$ is a $f,g$-gapped repeat, we have $j'\ge i'+c'+g(c')$,
thus, taking into account $i' \ge i-c'/6$ and $c'\ge c$, we obtain 
$j'\ge i+\frac{5}{6}c+g(c')$. Taking into account $c'\ge c$, by definition 
of $\partial^-_g$ we have $g(c)-g(c')\le\partial^-_g (c'-c)$. From $c\ge \frac{2c'}{3}$ 
we have also $c'-c\le c/2$. Thus $g(c')\ge g(c)-\partial^-_g\frac{c}{2}$, so
$$
j\ge j'\ge i+\frac{5}{6}c+g(c)-\partial^-_g\frac{c}{2}.
$$
Since $\sigma'$ is a $f,g$-gapped repeat, we have also $j'\le i'+c'+f(c')$,
so, taking into account $j \le j'+c'/6$, $c'\le\frac{3}{2}c$ and $i'\le i$, 
we obtain $j\le i+\frac{7}{4}c+f(c')$. Taking into account $0\le c'-c\le c/2$,
by definition of $\partial^+_f$ we have $f(c')-f(c)\le\partial^+_f (c'-c)\le
\partial^+_f\frac{c}{2}$. Thus
$$
j\le i+\frac{7}{4}c+f(c)+ \partial^+_f\frac{c}{2}.
$$

\begin{lemma}
Let both the values $\partial^-_f$, $\partial^+_g$ exist. Then for any point $(i, j, c)$ 
from ${\cal CQ}^b$ the inequalities
$$
i+\frac{5}{9}c+g(c)-\frac{c}{3}\partial^+_g\le j\le i+\frac{7}{6}c+f(c)+\frac{c}{3}\partial^-_f
$$
hold.
\label{coverlemmb}
\end{lemma}

{\bf Proof.} Let a point $(i, j, c)$ belongs to ${\cal CQ}^b$, i.e. this point
is covered from below by some point $(i', j', c')$ representing the gapped repeat
$\sigma'\equiv (u', u'')$ from ${\cal P}_{f, g}$ such that $i'={\rm beg} (u')$, 
$j'={\rm beg} (u'')$ and $c'=c(\sigma')$. Thus we have $i'\le i \le i'+c'/6$,
$j'\le j \le j'+c'/6$, and $c'\le c \le \frac{3c'}{2}$. Since $\sigma'$ is a 
$f,g$-gapped repeat, we have $j'\ge i'+c'+g(c')$, thus, taking into account 
$i'\ge i-c'/6$ and $c'\ge\frac{2}{3} c$, we obtain $j'\ge i+\frac{5}{9}c+g(c')$.
Taking into account $c\ge c'$, by definition of $\partial^+_g$ we have 
$g(c)-g(c')\le\partial^+_g (c-c')$. From $c'\ge\frac{2}{3} c$ we have also
$c-c'\le\frac{c}{3}$, so $g(c)-g(c')\le\partial^+_g\frac{c}{3}$. Thus
$$
j\ge j'\ge i+\frac{5}{9}c+g(c)-\partial^+_g\frac{c}{3}.
$$
We have also $j'\le i'+c'+f(c')$, so, taking into account $j \le j'+c'/6$, $c'\le c$
and $i'\le i$, we obtain $j\le i+\frac{7}{6}c+f(c')$. Taking into account 
$0\le c-c'\le c/3$, by definition of $\partial^-_f$ we have $f(c')-f(c)\le
\partial^-_f (c-c')\le\partial^-_f\frac{c}{3}$. Thus
$$
j\le i+\frac{7}{6}c+f(c)+ \partial^-_f\frac{c}{3}.
$$

Assign now to each point $(i, j, c)$ the weight $\rho (i, j, c)=1/c^3$,
and for any finite set~$A$ of points define 
$$
\rho (A)=\sum_{(i, j, c)\in A} \rho (i, j, c)=\sum_{(i, j, c)\in A}\frac{1}{c^3}.
$$
From Lemma~\ref{coverlemma} we obtain the following bound for $\rho ({\cal CQ}^a)$.

\begin{corollary}
Let the value $\partial^a_{f, g}$ exist. Then 
$\rho ({\cal CQ}^a)=O(n(1+\max\{\partial^a_{f, g}, \Delta_{f, g}\}))$.
\label{weightcorra}
\end{corollary}

{\bf Proof.} Since the value $\partial^a_{f, g}$ exist, 
the values $\partial^+_f$, $\partial^-_g$ are defined.
Denote $l(c, i)=i+\frac{5}{6}c+g(c)-\frac{c}{2}\partial^-_g$,
$u(c, i)=i+\frac{7}{4}c+f(c)+\frac{c}{2}\partial^+_f$. By Lemma~\ref{coverlemma}
each point $(i, j, c)$ from ${\cal CQ}^a$ satisfies the restrictions 
$l(c, i)\le j\le u(c, i)$. Thus
$$
\rho ({\cal CQ}^a)\le \sum_{c=1}^{n/2}\sum_{i=1}^{n}\sum_{l(c, i)\le j\le u(c, i)}\frac{1}{c^3}
\le \sum_{c=1}^{n/2}\sum_{i=1}^{n}\frac{u(c, i)-l(c, i)+1}{c^3}.
$$
Note that
\begin{eqnarray*}
\frac{u(c, i)-l(c, i)+1}{c^3}&=&\frac{1}{c^3}[1+\frac{11}{12}c+(f(c)-g(c))+
\frac{c}{2}(\partial^+_f+\partial^-_g)]\\
&<&\frac{1}{c^3}[1+c+c\Delta_{f, g}+c\partial^a_{f, g}]\le\frac{1}{c^2}[2+\Delta_{f, g}+\partial^a_{f, g}].
\end{eqnarray*}
Therefore
\begin{eqnarray*}
\rho ({\cal CQ}^a)&<&[2+\Delta_{f, g}+\partial^a_{f, g}]\sum_{c=1}^{n/2}\sum_{i=1}^{n}\frac{1}{c^2}<
[2+\Delta_{f, g}+\partial^a_{f, g}]n\sum_{c=1}^{\infty}\frac{1}{c^2}\\
&=&O(n(1+\max\{\partial^a_{f, g}, \Delta_{f, g}\})).
\end{eqnarray*}

An analogous bound for $\rho ({\cal CQ}^b)$ can be obtained from Lemma~\ref{coverlemmb}.

\begin{corollary}
Let the value $\partial^b_{f, g}$ exist. Then 
$\rho ({\cal CQ}^b)=O(n(1+\max\{\partial^b_{f, g}, \Delta_{f, g}\}))$.
\label{weightcorrb}
\end{corollary}

\begin{lemma}
Let the values $\partial^a_{f, g}$ exist. Then 
$|{\cal OP}_{f, g}|=O(n(1+\max\{\partial^a_{f, g}, \Delta_{f, g}\}))$.
\label{OPlemma}
\end{lemma}

{\bf Proof.} Let $\sigma$ be an arbitrary repeat from ${\cal OP}_{f, g}$ represented by a point 
$(i', j', c')$. Then
\begin{eqnarray*}
\rho (V[\sigma ])&=&\sum_{i'\le i\le i'+c'/6}\;\sum_{j'\le j\le j'+c'/6}\;\sum_{2c'/3\le c\le c'}\frac{1}{c^3}\\
&>&\frac{c'^2}{36}\sum_{2c'/3\le c\le c'}\frac{1}{c^3}.
\end{eqnarray*}
Using the standard estimation for sums with integrals, one can show that 
$\sum_{2c'/3\le c\le c'}\frac{1}{c^3}\ge \frac{5}{32}\frac{1}{c'^2}$ for any~$c'$.
Thus for any $\sigma$ from ${\cal OP}_{f, g}$
$$
\rho (V[\sigma ])>\frac{c'^2}{36}\frac{5}{32}\frac{1}{c'^2}=\Omega (1).
$$
Therefore,
\begin{equation}
\sum_{\sigma\in {\cal OP}_{f, g}}\rho (V[\sigma ])=\Omega (|{\cal OP}_{f, g}|).
\label{lowbndOP}
\end{equation}
Note that any point covered by repeats from ${\cal OP}_{f, g}$ belongs to ${\cal CQ}_a$.
On the other hand, by Lemma~\ref{coverlemma}, each point of ${\cal CQ}_a$ can not be covered 
by two repeats from ${\cal OP}_{f, g}$. Thus
$$
\sum_{\sigma\in {\cal OP}_{f, g}}\rho (V[\sigma ])\le \rho ({\cal CQ}_a),
$$
so by Corollary~\ref{weightcorra}
$$
\sum_{\sigma\in {\cal OP}_{f, g}}\rho (V[\sigma ])=O(n(1+\max\{\partial^a_{f, g}, \Delta_{f, g}\})).
$$
This relation together with the relation~\ref{lowbndOP} implies 
$$
|{\cal OP}_{f, g}|=O(n(1+\max\{\partial^a_{f, g}, \Delta_{f, g}\})).
$$

Using Lemma~\ref{coverlemmb} and Corollary~\ref{weightcorrb}, analogously to Lemma~\ref{OPlemma} we can 
obtain also the following bound for $|{\cal OP}_{f, g}|$.
\begin{lemma}
Let the value $\partial^b_{f, g}$ exist. Then
$|{\cal OP}_{f, g}|=O(n(1+\max\{\partial^b_{f, g}, \Delta_{f, g}\}))$.
\label{OPlemmb}
\end{lemma}

We combine Lemma~\ref{OPlemma} and Lemma~\ref{OPlemmb} together for estimating $|{\cal OP}_{f, g}|$.
\begin{corollary}
$|{\cal OP}_{f, g}|=O(n(1+\max\{\partial_{f, g}, \Delta_{f, g}\}))$.
\label{OPcorr}
\end{corollary}

Summing up Corollaries~\ref{onPP}, \ref{onSP} and~\ref{OPcorr} together, we obtain the following
upper bound on the number of maximal $f,g$-gapped repeats in~$w$.
\begin{theorem}
Let both the values $\partial_{f, g}$, $\Delta_{f, g}$ exist. Then a word~$w$ of length~$n$
contains $O(n(1+\max\{\partial_{f, g}, \Delta_{f, g}\}))$ maximal $f,g$-gapped repeats.
\label{repsbound}
\end{theorem}

\section{Conclusion}

Recall that $\alpha$-gapped repeats can be considered as a particular case of $f,g$-gapped 
repeats for $f(x)=(\alpha -1)x$ and $g(x)=\min \{1, \alpha -1\}$, so the obtained 
$O(n(1+\max\{\partial_{f, g}, \Delta_{f, g}\}))$ bound on the number of maximal $f,g$-gapped 
repeats implies $O(\alpha n)$ bound obtained in~\cite{LATA16} on the number of maximal 
$\alpha$-gapped repeats. Note that in the paper we don't concern algorithmic questions of
finding of maximal $f,g$-gapped repeats in words. However it would be interesting if there exists
an algorithm for finding of all maximal $f,g$-gapped repeats in a word with the same 
$O(n(1+\max\{\partial_{f, g}, \Delta_{f, g}\}))$ time bound. Note also that even if
the obtained bound is tight for some cases of words a further strengthening of this bound
could be interesting.

\end{document}